\documentclass{aastex631}

\usepackage{amsmath}

\newcommand{\lyal}{Lyman-$\alpha$ }

\newcommand{\hi}{H$\,$\textsc{i} }
\newcommand{\his}{H$\,$\textsc{i}}

\newcommand{\mathd}{\mathrm{d}}

\newcommand{\pow}[1]{\times 10^{#1}}

\begin{document}

\title[]{21~cm Signal from the Thermal Evolution of \lyal during Cosmic Dawn}

\author[0000-0001-7451-6139]{Janakee Raste}
\affiliation{National Centre for Radio Astrophysics, 
Tata Institute of Fundamental Research, Pune 411007, India}

\author{Shiv K. Sethi}
\affiliation{Raman Research Institute, Bengaluru 560080, India}

\correspondingauthor{Janakee Raste}
\email{janakee@ncra.tifr.res.in}

\begin{abstract}

The \lyal photons couple the spin temperature of neutral hydrogen (\his) to the kinetic temperature during the era of cosmic dawn. During this process, they also exchange energy with the medium, heating and cooling the \his. In addition, we expect X-ray photons to heat the mostly neutral gas during this era. We solve this coupled system (Lyman-$\alpha$--\hi system along with X-ray heating) for a period of 500~Myr (redshift range $8 <z < 25$). Our main results are: (a) Without X-ray heating, the temperature of the gas reaches an equilibrium, which is nearly independent of photon intensity and only weakly dependent on the expansion of the Universe. The main determinant of the quasi-static temperature is the ratio of injected and continuum Lyman-$\alpha$ photons. (b) While X-ray photons provide an additional source of heating at initial times, for large enough Lyman-$\alpha$ photon intensity, the system tends to reach the same  quasi-static temperature as expected without additional heating. This limit is reached when the density of photons close to the  Lyman-$\alpha$ resonance  far exceeds  the \hi number density. (c) We compute the global \hi signal for these scenarios. In the limit of the large density of Lyman-$\alpha$ photons, the spin temperature of the hyperfine line is fixed. This freezes the global \hi signal from the era of cosmic dawn and the crossover redshift from absorption to emission. This feature depends only on the ratio of injected to continuum Lyman-$\alpha$ photons, and the global \hi signal can help us determine this ratio. 
\end{abstract}

\section{Introduction} \label{sec:intro}

\hi cosmology has emerged as an important area of cosmology research in the past two decades. Its main appeal arises from the
fact that the redshifted \hi line can be used to study three important eras---dark ages ($200 \lesssim z  \lesssim 35$), cosmic dawn (CD; $35 \lesssim z  \lesssim 15$), and Epoch of Reionization (EoR; $15 \lesssim z  \lesssim 6$)---which are not easily accessible to other cosmological
probes.  Our current partial understanding of these eras is based on the Gunn-Peterson (GP) test on high-redshift quasars at $z\simeq 6$ \citep{Fan:2000gq,2001AJ....122.2850B} and the detection of cosmic microwave background (CMB) temperature and polarization anisotropies from WMAP and Planck satellite missions \citep{Hinshaw:2012aka,Planck2018}. Most recent Planck results determine the epoch of reionization, $z_{\rm reion} = 7.75 \pm 0.73$.
These probes suggest that the Universe might have made a transition from fully neutral to highly ionized in the
redshift range $6 <z < 10$. These results are supplemented by recent \lyal forest observations, which suggest that the EoR might have ended at $z\simeq 5.3$ \citep{2023ApJ...955..115Z, 2021MNRAS.508.1853B, 2022MNRAS.514...55B, 2019MNRAS.485L..24K, RK21}.

The current theoretical paradigm, based on the $\Lambda$CDM model, suggests that the dark ages ended around $z \simeq 30$ with the formation
of the first objects in the Universe. These subgalactic structures emitted UV and X-ray radiation into the medium, which heated and ionized the gas, while also coupling the Lyman-$\alpha$ radiation to the \hi level population. For the redshift range, $10 \lesssim z \lesssim 30$, the global ionized fraction is small, but it increases rapidly for smaller redshifts, ending the EoR at $z\simeq 6$ \citep{Barkana:2000fd,21cm_21cen,2014PTEP.2014fB112N,2010ARA&A..48..127M}. Although this paradigm is generally accepted, 
we currently lack detailed information on the physics of the first stars and galaxies, a situation that has become even more
complex with JWST results \citep{yan2022first, hassan2023jwst}.

The redshifted 21~cm  signal from  the era of CD/EoR is determined by three radiation fields: hydrogen-ionizing photons, photons in the frequency range between Lyman-$\alpha$ and Lyman-limit (referred to as `Lyman-$\alpha$' radiation, this radiation  couples Lyman-$\alpha$ photons  to \hi hyperfine levels through the Wouthuysen-Field effect), and X-ray radiation (photons in the approximate energy range 100~eV--10~keV; the main role of these photons is to heat the neutral gas; e.g., \cite{RS18, RS19,Sethi05,2006PhR...433..181F,21cm_21cen} and references therein).

In this paper, we investigate the thermal impact of Lyman-$\alpha$ photons. This is a less studied aspect of Lyman-$\alpha$ photons
because the number density of Lyman-$\alpha$ needed for Wouthuysen-Field coupling is too small to have significant thermal impact (\cite{2004ApJ...602....1C};  see also \cite{1997ApJ...475..429M}).
This claim has since been confirmed by other authors (e.g., \cite{chuzhoy2006ultraviolet,chuzhoy2007heating,rybicki2006improved,meiksin2006energy,hirata2006wouthuysen,2006MNRAS.367.1057P}; see also \cite{mittal2021ly, mittal2023radiative,munoz2022impact,ghara2020impact,semelin2023accurate,meiksin2021impact,shimabukuro2023exploring,venumadhav2018heating,2021MNRAS.506.5479R}). In \cite{2024ApJ...976..236R} (hereafter Paper~I), we studied  this issue as an initial value problem and solved the simultaneous evolution 
of Lyman-$\alpha$ spectral profile and the gas temperature. Many initial profiles (injected, continuum, and a mix of the two) were
considered. It was shown that both the spectral profile and the gas temperature reach quasi-equilibrium on time scales that
depend on the photon injection rate. For large photon injection rates, this time scale is less than a few million years. While Paper~I focused on a detailed theoretical analysis, our aim in this paper is to extend this analysis to enable us to compute cosmological observables. 

In Paper~I, the coupled photon-gas system was evolved to a few million years, but this only allowed us to understand the
equilibrium states for large photon injection rates. In this paper, we evolve the system to nearly half a billion years, which
enables us to treat small photon injection rates adequately, clearly discern the effect of expansion of the Universe, and reliably compute the cosmological observables. Second, 
we include  X-ray heating  as an additional source in our work. In the usual analysis, Lyman-$\alpha$ and X-ray photons
are treated as uncoupled. However, our analysis shows that they are strongly coupled for a large Lyman-$\alpha$ injection rate. Third,
this enables us to compute the global \hi signal during the CD/EoR eras. In particular, our analysis reveals the striking impact of Lyman-$\alpha$ and X-ray coupling on the observed \hi signal. 

In the next section,
we discuss our formalism in brief and defer to Paper~I for a more detailed discussion.   In Section~\ref{sec:results}, we show
the long-term evolution of the photon profile and gas temperature. In Section~\ref{sec:xrayheat} we study the implications
of X-ray heating within the framework of many models of Lyman-$\alpha$ and X-ray injection. The global \hi signal is also computed
for these models. Finally, in Section~\ref{sec:conc}, 
we summarize our results and provide possible future avenues of our work.  Throughout the paper, we assume the best-fit Planck
parameters corresponding to the spatially flat FRW Universe: $\Omega_c h^2 = 0.12$ and $\Omega_b h^2 = 0.0224$ \citep{Planck2018}.

\section{Thermal coupling of \texorpdfstring{Lyman-$\alpha$}{Lyman-alpha} photons with \his: X-ray heating } \label{sec:lyah1}
In paper~I, we developed the formalism to determine the thermal coupling of Lyman-$\alpha$ photons and neutral hydrogen. We give here only the equations
directly relevant for this work and refer the reader to Paper~I for more details.

Our aim is to solve for the evolution of the photon occupation number in an FRW Universe. Let $J(\hat n,\nu,t) \equiv dN/(d\nu dV d\Omega)$ denote the number density of photons in the frequency range  $\nu$ and $\nu+d\nu$, travelling in
a direction $\hat n$. It is more convenient to work with variable $x= (\nu-\nu_\alpha)/\Delta\nu_D$ where  $\Delta\nu_D = \nu_\alpha v_T/c$
is the Doppler width with thermal velocity $v_T = \sqrt{2k T_{\rm K}/m_p}$. This allows us to define $J(x,t) = J(\nu,t) \Delta\nu_D$.
$J(x,t)$ is the number density of photons per unit solid angle in the frequency range corresponding to $x$ and $x+dx$. The evolution of 
$J(x,t)$ can be written as (for details, see Paper~I and references therein): 
\begin{equation}
  {1\over a^3}{\partial (J(x,t) a^3) \over \partial t'}    = {\partial \over \partial x} \left [{\phi(x) \over 2} {\partial J(x,t) \over \partial x} +\left(\eta\phi(x) + \gamma \right) J(x,t) \right ] + C' \psi(x) .
  \label{eq:fineq1}
\end{equation}
Here, the dimensionless time is defined as $t' = t/t_{\rm sca}$, where $t_{\rm sca} =  1/(n \sigma_0 c(1+w/T_{\rm K}))$ is the the scattering time at the line center.
The parameter $w =  b \nu_{21}^2 m_p c^2/(2\nu_\alpha^2 k)$, which captures the contribution of hyperfine energy exchange, is small ($w \simeq 0.4 \, \rm K$), and the impact of hyperfine energy exchange is negligible except at very low temperatures. $\sigma_0 = (3/8\pi) (\lambda_\alpha^2 A_\alpha/\Delta\nu_D)$ is the Lyman-$\alpha$ scattering cross section at the line center. The Sobolev parameter $\gamma = \tau_{\rm GP}^{-1} (1+w/T_{\rm K})^{-1}$ compares the relative efficacy of scattering and expansion, with the GP optical depth $\tau_{\rm GP} =\sigma_0 n v_T/H$. The recoil parameter $\eta = (1+w/T_{\rm S})(1+w/T_{\rm K})^{-1} (h\nu_\alpha/(2kT_{\rm K}m_pc^2)^{1/2})$. $C'$ is the rescaled photon injection rate: $C' = C/(n\sigma_0 c (1+w/T_{\rm K}))$; $C$ is the rate at which the new photons are injected. For our work, we define $C$ as the number density of new photons that are produced per unit time per unit solid angle in the frequency range from Lyman-$\alpha$ to Lyman limit (in $ \rm cm^{-3} \, sec^{-1} \, sr^{-1}$). $\psi(x)$, the normalized photon injection profile, is defined such that $\int \psi(x) dx = 1$. 

To compute relevant variables, we need the spin temperature, $T_{\rm S}$, which can be expressed as: 
\begin{equation}
  T_{\rm S} = {T_{\rm CMB} + y_c T_{\rm K} + y_\alpha T_\alpha \over 1+ y_c + y_\alpha},
  \label{eq:ts_def}
\end{equation}
and $T_\alpha$, the colour temperature close to the resonance line is,
\begin{align}
    T_\alpha = - {h(\nu - \nu_\alpha) \over k\; \; \ln\left( J(\nu) \over J(\nu_\alpha) \right)}.
    \label{eq:tadef}
\end{align}
The \lyal coupling coefficient,
\begin{equation}
y_\alpha = (h\nu_{21}/kT_\alpha)(P_{21}/A_{21}),
\label{eq:yalpha}
\end{equation}
with  the rate of deexcitation of the upper hyperfine level, $P_{21} = (4/27) P_\alpha$ and  $P_\alpha \simeq  4\pi \sigma_0 J(\nu_\alpha) c$
(see e.g., \cite{2010ARA&A..48..127M, field1958excitation} for more details). In thermal equilibrium between Lyman-$\alpha$ photons and the intergalactic medium (IGM), $T_\alpha = T_{\rm K}$. 
If $y_\alpha \gg y_c$ and $y_\alpha \gg 1$, then $T_{\rm S} = T_\alpha$. 

We further redefine $J(x,t) \to a^3 J(x,t)$ and $C' \to C' a^3$, converting the photon intensity and the photon production rate into comoving quantities. This gives us,
\begin{equation}
  {\partial J(x,t) \over \partial t'}    = {\partial \over \partial x} \left [{\phi(x) \over 2} {\partial J(x,t) \over \partial x} +\left(\eta\phi(x) + \gamma \right) J(x,t) \right ] + C' \psi(x).
  \label{eq:fineq2}
\end{equation}
This equation is similar to the one derived by \citep{1994ApJ...427..603R} with two key differences: it includes a factor of $a^3$ instead of $a^2$, and it correctly accounts for energy exchange owing to hyperfine mixing. It can be readily shown that $C't' = Ct$ is the photon number density integrated over all frequencies, which is consistent with the conservation of the number of photons.  

The evolution of $J(x)$ and the spin temperature, $T_{\rm S}$ is determined by the thermal state of the gas. The evolution of  kinetic temperature $T_{\rm K}$ for a neutral, monoatomic gas, is given by:
\begin{equation}
{dT_{\rm K} \over dt} = -2 H T_{\rm K} + {2\over 3} {(\dot q_\alpha + \dot q_{\rm xray}) \over n_b k}.
\label{eq:thermal}
  \end{equation}
Here, $\dot q_{\rm xray}$ is the energy injection owing to X-ray photons and $\dot q_\alpha = \dot Q n_b  $ is the rate at which the energy is injected  by Lyman-$\alpha$ photons per unit volume ($\rm erg \, cm^{-3} \, s^{-1}$; for details see Paper~I), with: 
\begin{eqnarray}
    \dot Q &=&  4\pi c\int  {(h\nu_\alpha)^2 \over m_p c^2} \sigma_0\phi(x)  \left(J(x,t) + {kT_{\rm K} \over \Delta \nu_D h} J'(x,t) \right) dx\nonumber \\
    &+&  4 \pi c\int  {b(h\nu_{21})^2 \over 2k T_{\rm S}} \sigma_0\phi(x)  \left(J(x,t) + {kT_{\rm S} \over \Delta \nu_D h} J'(x,t) \right) dx,
    \label{eq:ene_exchange1}
\end{eqnarray}
and $J'(x,t) = \mathd J/\mathd x$.

Simultaneous solutions to Eqs.~(\ref{eq:ts_def}), (\ref{eq:fineq2}), (\ref{eq:thermal}), and~(\ref{eq:ene_exchange1})  
yield the evolution of the photon profile along with the hyperfine and thermal state of \his. 

\section{Long-term evolution} \label{sec:results}

\begin{figure*}
    \includegraphics[width=1\linewidth]{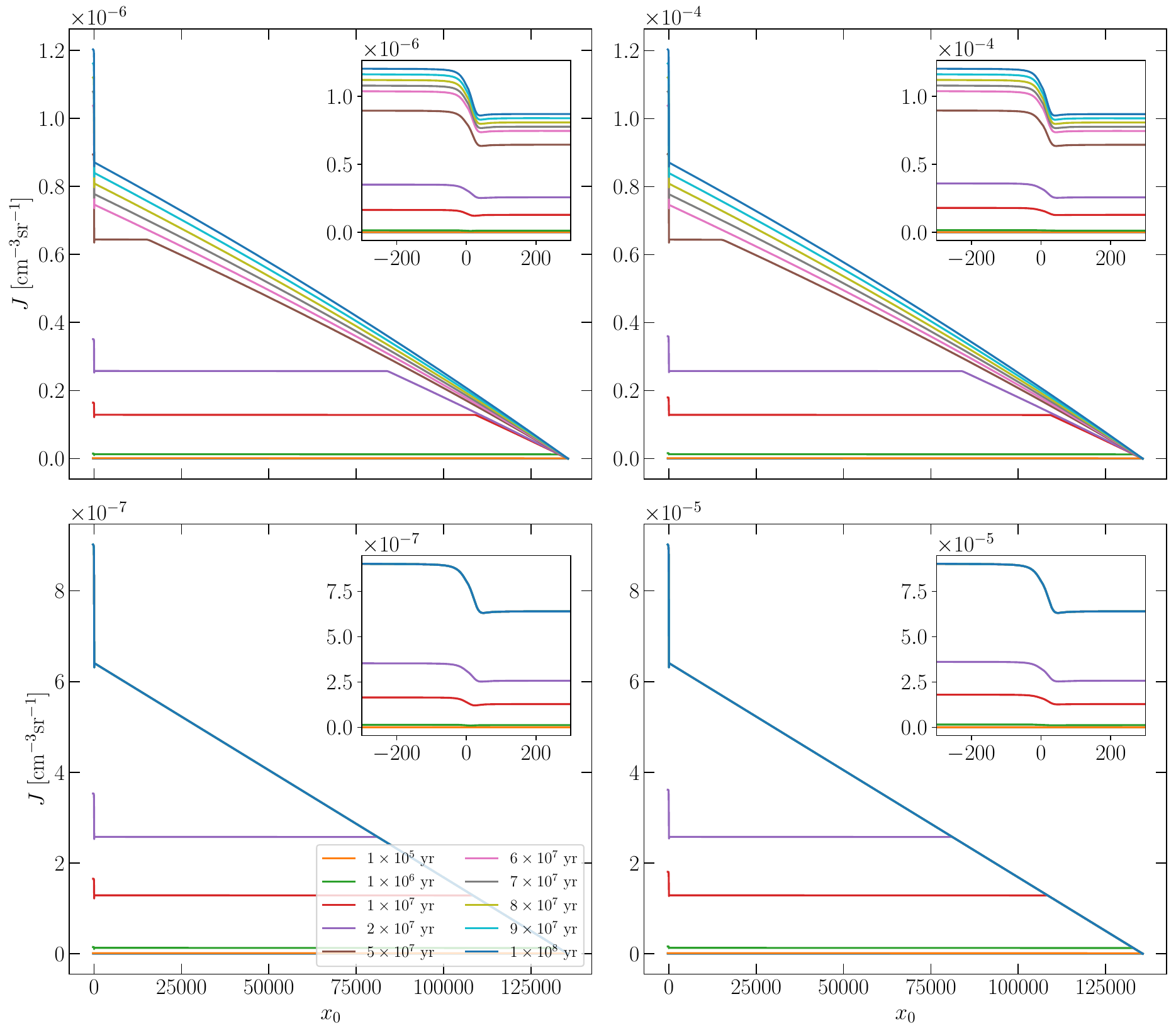} 
  \caption{We show the evolution of intensity profiles for \lyal continuum + 20\% injected photons for 100~Myr. The left (right) panels correspond to $C=1\pow{-16}$ $ \rm cm^{-3} \, s^{-1} \, sr^{-1}$ ($C=1\pow{-14}$ $ \rm cm^{-3} \, s^{-1} \, sr^{-1}$). The following initial conditions are used: $z = 20$, $T = 10 \, \rm K$, fully neutral \hi gas; the grid on the x-axis corresponds to the variable: $x_0=x(T=10\, \rm K)$. The evolution of the profile close to  $x=0$ is shown in the inset. The bottom panels assume local approximation. In this approximation, the expansion rate and the number density of \hi atoms are  held fixed to their  initial value  for solving Eqs.~(\ref{eq:ts_def}), (\ref{eq:fineq2}), (\ref{eq:thermal}), and~(\ref{eq:ene_exchange1}). If this evolution is switched off, a quasi-static equilibrium is reached after nearly 50~Myr. The top panels, the realistic case in which all relevant quantities are allowed to evolve, show slow change of the photon profile even  at late times.}
  \label{fig:long_noz}
\end{figure*}

\begin{figure*}
  \includegraphics[width=1.0\linewidth]{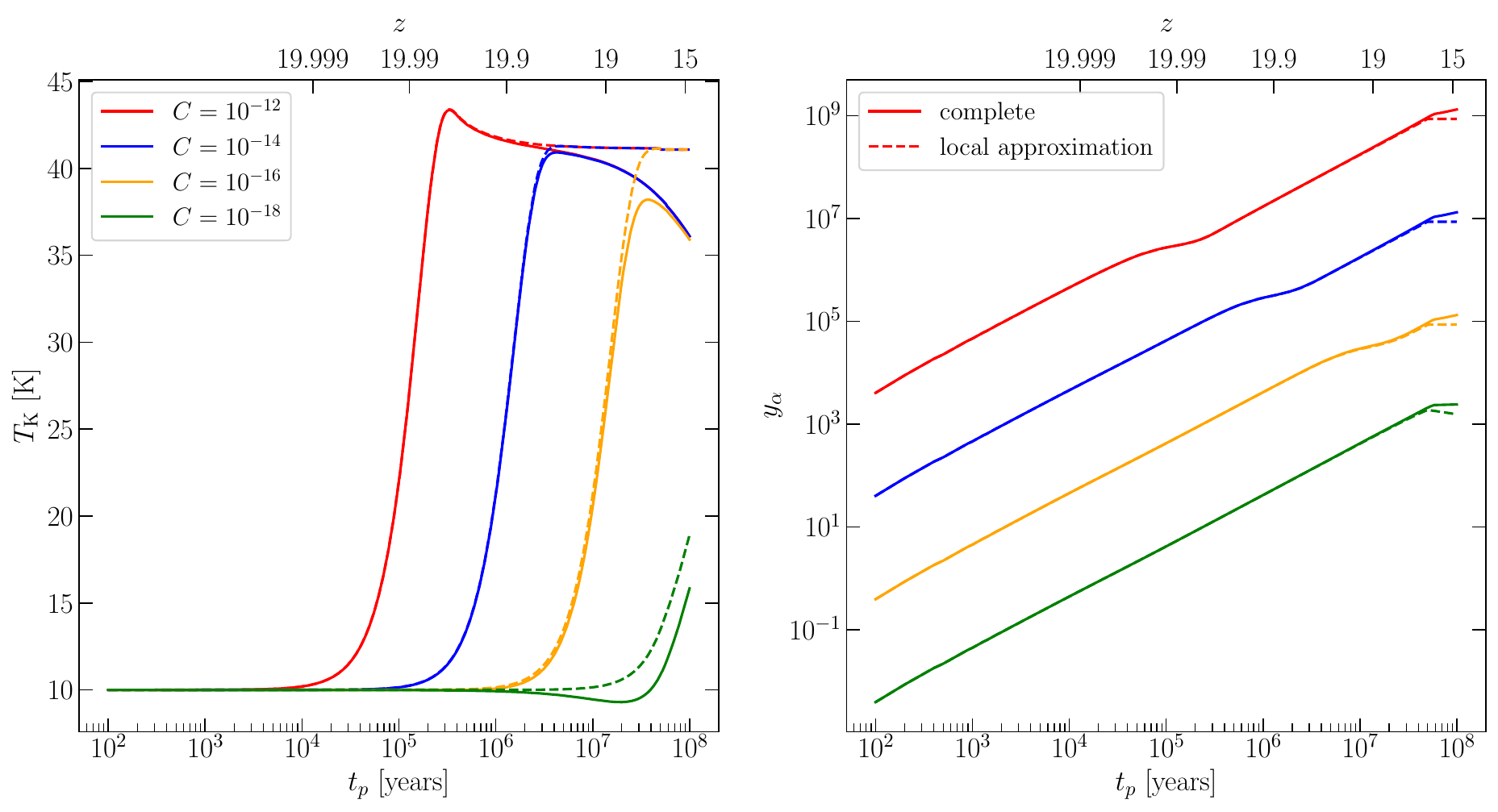}
  \caption{In the left panel, we show the thermal evolution of the gas for the photon injection model (\lyal continuum + 20\% injected photons) and initial conditions  as in  Figure~\ref{fig:long_noz}   for 100~Myr with different photon injection rates, $C$.  The solid curves include all the relevant physics, while for the dashed curves correspond to the aforementioned local approximation (Figure~\ref{fig:long_noz}). In local approximation, a quasi-static thermal equilibrium is  reached, while a slow evolution of the gas temperature is seen in the other case. For a very small value of $C$, the temperature initially decreases due to adiabatic cooling  before the \lyal intensity builds up.  In the right panel, we show the evolution of the coupling coefficient $y_\alpha$ for the same cases. $y_\alpha$ is proportional to $J(\nu_\alpha)$ and inversely proportional to $T_\alpha$ (Eq.~(\ref{eq:yalpha})). Initially, $y_\alpha$ increases rapidly as $J(\nu_\alpha)$ (or equivalently $J(x=0)$) increases  rapidly (Figure~\ref{fig:long_noz}). If the temperature also increases sharply (left panel), $y_\alpha$ flattens. Over longer times, after the tilted profile is reached, $J(0)$ is nearly constant, but the temperature decreases slightly, leading to a small increase in $y_\alpha$. In  local approximation, $y_\alpha$ reaches a constant as both $J(0)$ and the temperature reach steady state values (left panel and Figure~\ref{fig:long_noz}).}
  \label{fig:long_ns}
\end{figure*}

In Paper I, we studied the evolution of intensity profiles for short periods $t \lesssim  5 \, \rm Myrs$. This time scale suffices for many purposes, e.g., for establishing an approximate equilibrium gas temperature. However, we also showed that photon profiles
in some cases (e.g., the continuum and continuum-plus-injected intensity profiles) display long-term evolution. We dealt with these cases by employing a smaller (and unrealistic) values of $x_{\rm max}$. In this work, we explore the more realistic case, $x_{\rm max} \simeq 10^5$.  This requires us to evolve the system for more than 100~million years. The long-term evolution also enables us
to study the approach to equilibrium for smaller photon injection rates, $C$, than was possible in our earlier work. 

In Figure~\ref{fig:long_noz}, we show the long-term evolution of intensity profiles for  $x_{\rm max} = 1.4\times 10^5$, which corresponds to the frequency range between Lyman-$\alpha$ and Lyman-$\beta$. To illustrate approach to equilibrium profile, one can consider many cases: (a) neglect the expansion of the Universe. Without expansion, the equilibrium profile $\propto \exp(-h(\nu-\nu_\alpha)/(kT))$ (for details, see Paper~I and \cite{rybicki1994accelerated,chugai1980scattering,chuzhoy2006ultraviolet}), (b) neglect scattering. This case was discussed in detail in Paper~I. This allows us to compute the relevant time scales for reaching equilibrium analytically. As we showed 
in Paper~I, these solutions guide us in understanding the equilibrium in realistic cases. In this paper, 
we consider another approximation to study the dynamics of the coupled  system: we include both the scattering and expansion mechanisms but assume a constant expansion rate and no evolution of other relevant quantities, such as number density. This approximation allows us to 
separate the impact of long-term evolution from the short-term evolution of the system. We evolve the system for close to 100~million years, a period over which both the number density and the expansion rate change substantially.  However, by holding these quantities constant, as we show below, we can better understand the quasi-equilibrium states that are reached in shorter times. Therefore, the aim of this approximation is to isolate the impact of different physical processes. We refer to this case as ``local approximation". 

Figure~\ref{fig:long_noz} (lower panels)  shows that for the local approximation, a nonevolving intensity profile is reached in nearly 50~million years.  In Paper~I, we analytically derived the evolution of photon spectra for both injected and continuum profiles for the no-scattering case and argued they 
give a reasonable approximation of the time-scale over which the equilibrium is reached and equilibrium intensity profiles in most realistic cases. Although it is
not possible to analytically derive the photon spectra when both injected and continuum photons are present, the equilibrium profiles displayed in the lower panels of Figure~\ref{fig:long_noz} are consistent with our general findings of Paper~I. In particular,
we notice how the flat profile makes a transition to a quasi-static    tilted profile and the tilt moves to smaller frequency with time. For us,
the relevant time scale is the time over which the profile around $x\simeq 0$ becomes tilted; this time scale corresponds to the time over which the photons redshift from $x = x_{\rm max}$ to $x = 0$. 

In the upper panels
of Figure~\ref{fig:long_noz}, we show the long-term evolution of the spectral profile when all the relevant physics is included. In this
case, we notice that the photon profiles do not relax to quasi-steady state but keep evolving even after the profile becomes tilted. A comparison with the lower panels based on local approximation shows the main
cause of this evolution to be the redshift evolution of the expansion rate and the number density of \hi atoms, which becomes important over the long-term evolution of the system.  We notice that
we did not see this behaviour in Paper~I because the evolution time scale was much shorter. 

In Figure~\ref{fig:long_ns}, we show the long-term temperature evolution for a range of models. Except for the smallest photon rate,
the temperature reaches an equilibrium value in $t \lesssim 100 \, \rm Myr$.  In the local approximation (case (c) above), the temperature 
reaches the same constant value independent of the photon injection rate, $C$; this was anticipated in Paper~I based on shorter time scales\footnote{In paper~I, using on a smaller, unrealistic value of $x_{\rm max}$, we also argued that when  the tilted profile is reached at $x = 0$, the equilibrium temperature decreases by a small amount. Here, we verify this result on the basis of long-term thermal evolution.  In Figure~\ref{fig:long_ns}, we notice a small step function in temperature in dashed lines close to 50~Myr.}. The thermal evolution for the local approximation is consistent with the corresponding photon profiles in Figure~\ref{fig:long_noz}, as both the photon profile and the gas temperature reach quasi-steady states. 
The redshift evolution of relevant quantities causes both the IGM temperature and the intensity
profile to slowly evolve. However, Figure~\ref{fig:long_ns} shows that the local approximation gives a fairly precise value of the equilibrium temperature. 

In the right panel of Figure~\ref{fig:long_ns}, the evolution of $y_\alpha$, the parameter that denotes
the efficiency of coupling between the spin temperature, $T_{\rm S}$, of the 21~cm line and Lyman-$\alpha$ colour temperature, $T_\alpha$, is shown. The dimensionless parameter $y_\alpha$ increases with the Lyman-$\alpha$ intensity and is inversely proportional to 
the colour temperature, $T_\alpha$ (Eq.~(\ref{eq:yalpha})).  As $T_\alpha = T_{\rm K}$, the  evolution of $y_\alpha$  can be understood from the left panel of Figure~\ref{fig:long_ns} and Figure~\ref{fig:long_noz}. For $y_\alpha >  1$, $T_{\rm S} \simeq T_\alpha$. The right panel of Figure~\ref{fig:long_ns} shows that this coupling becomes efficient over time scales shorter than the expansion time scale for
all models we consider. We return to the implications of this in a later section.

\section{Impact of X-ray heating} \label{sec:xrayheat}

\begin{figure*}
  \includegraphics[width=1\linewidth]{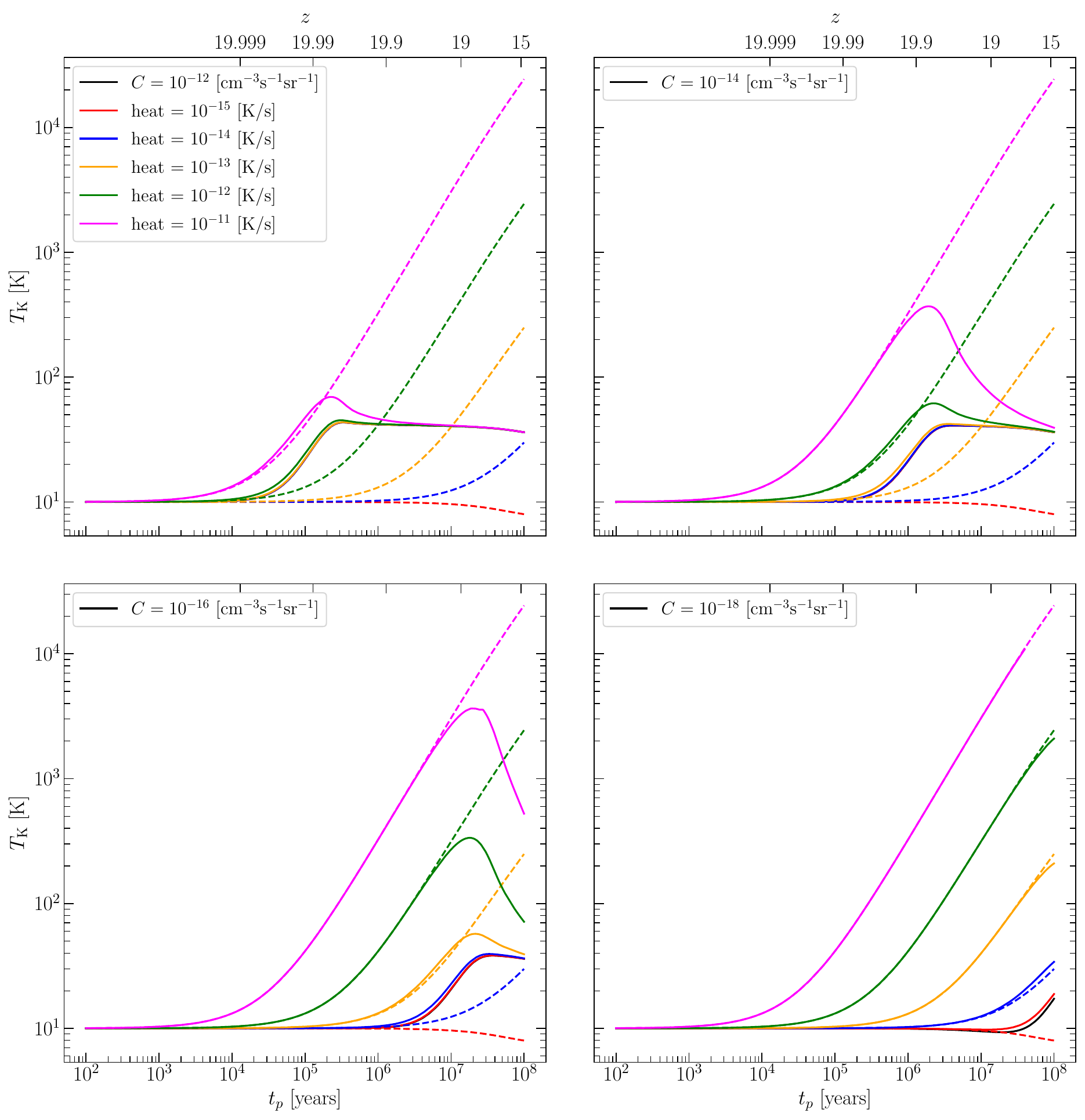}
  \caption{We show the temperature evolution of the gas including the impact of X-ray heating. The solid lines (dashed lines) correspond to cases when the thermal  effect of \lyal photons is included (excluded). The model considered is the same as in Figure~\ref{fig:long_noz}: continuum + 20\% injected Lyman-$\alpha$ photons with initial gas temperature, $T = 10 \, \rm K$ at $z = 20$. In each panel, the excess heating is varied over four orders of magnitude, while the \lyal injection rates are:  $C = 10^{-12}$ $ \rm cm^{-3} \, s^{-1} \, sr^{-1}$ (top left), $10^{-14}$ $ \rm cm^{-3} \, s^{-1} \, sr^{-1}$ (top right), $10^{-16}$ $ \rm cm^{-3} \, s^{-1} \, sr^{-1}$ (bottom left), and $10^{-18}$ $ \rm cm^{-3} \, s^{-1} \, sr^{-1}$ (bottom right). At initial times, both X-ray and Lyman-$\alpha$ photons act as heating sources. However, the net impact of Lyman-$\alpha$ photons is to drive the system toward an equilibrium temperature, in which the gas  and Lyman-$\alpha$ colour temperature approach each other. This causes  a fraction of the additional heat to be shared with Lyman-$\alpha$ photons, thereby preventing the X-ray photons  from heating the medium. For most models shown in the four panels, this behaviour is seen, with the equilibrium reaching faster for larger $C$. Only for $C = 10^{-18}$ the impact of Lyman-$\alpha$ photons is negligible.}
  \label{fig:heat}
\end{figure*}

During the eras of CD/EoR, the neutral gas is also heated by X-ray photons emitted from ionizing sources. Here,
we include the impact of this additional source of heating.  In this section, our aim is to discern generic outcomes, so we do not
consider specific models based on CD/EoR physics. We consider such models in the next section. 

In Figure~\ref{fig:heat}, we show the thermal evolution of the \hi gas in the presence of \lyal photons along with another heating source. 
To capture the full impact of the interaction of the new heating source with the coupled Lyman-$\alpha$-\hi system, we consider
four orders of magnitude variation in the heating rate of the additional source and six orders of magnitude variation in the Lyman-$\alpha$ photon injection rates in Figure~\ref{fig:heat}. 

The following inferences  can be drawn from Figure~\ref{fig:heat}: (a) For 
the Lyman-$\alpha$ injection rates  lower than $C \lesssim 10^{-18}$, Lyman-$\alpha$ photons have a negligible impact on X-ray heating (bottom right panel). (b) For large Lyman-$\alpha$ injection rates, $C \gtrsim 10^{-12}$, the thermal evolution  of the neutral gas is largely determined 
by Lyman-$\alpha$ photons (the top left panel of Figure~\ref{fig:heat} should be compared  to the left panel of Figure~\ref{fig:long_ns}). 
(c) The main tendency of Lyman-$\alpha$ photons is to drive the coupled system to equilibrium temperature,  which is nearly  independent of the photon injection rate, $C$ (Figure~\ref{fig:long_ns}). (d) At early times, a mix of Lyman-$\alpha$ and X-ray photons  can cause additional heating. However, the long-term evolution is determined by Lyman-$\alpha$ photons, preventing X-ray photons from heating the IGM.

Figure~\ref{fig:heat} shows that Lyman-$\alpha$ photons act as a sink for the additional source of heating. To understand this behaviour, we briefly review the case without this additional heating. For any initial condition, the coupled system of Lyman-$\alpha$ photons and \hi tends to reach an equilibrium, which drives the gas temperature and Lyman-$\alpha$ colour temperature (close to the line center $x\simeq 0$) toward each other. For continuum photons, the initial colour temperature is infinite, while it is zero for injected
photons.  This is the reason continuum photons heat while the injected photons cool.  For a mix of continuum and injected photons, with the former providing the dominant contribution, we get initial heating followed by an approach to equilibrium.  

We next try to determine the details of the dynamics of the coupled system when an additional source of heating of the gas is switched on: 

\begin{itemize}
\item[(a)] First, we qualitatively discuss what we expect when the  Lyman-$\alpha$ photons and the neutral gas are already in thermal equilibrium with each other, owing to Compton and inverse Compton scattering close to the line center. In this case, if either of the components receives additional energy, it would be shared with the other as the system attempts to reach a new equilibrium state.  For instance,  if continuum photons are added, the gas would heat as these photons  have an excess of blueward photons over the equilibrium state. Similarly, the addition of injected photons, which have an excess of redward photons, causes the gas to cool as energy is transferred from the gas to photons in reaching the new equilibrium (for details see Paper~I). For the same reason, if  the gas is heated (e.g. by X-ray photons), a fraction of the additional energy is absorbed by Lyman-$\alpha$ photons. The magnitude of this fraction is determined by the relative heat capacities of the two components. The results shown in Figure~\ref{fig:heat} are compatible with this qualitative understanding.  However, for the general case we consider, it is not possible to gain analytic insight into this process. Therefore, we consider a simpler case---no expansion and  no photon pumping ($H = C = 0$)---to get a clearer understanding. 
\item[(b)]  For the simpler case,  as shown in Paper~I, Compton and the inverse Compton scattering relax the coupled system to a thermal equilibrium with a common temperature $T_{\rm K}$ and the spectrum of the radiation in equilibrium  is $J(x) \propto \exp(2\eta x) = \exp(-h (\nu- \nu_\alpha)/kT_{\rm K})$ \footnote{Clearly, this  spectrum cannot
capture the entire picture, as it is divergent for $\nu \ll \nu_\alpha$. We discuss solutions far away from the line center in Paper~I.  However, this solution provides an excellent approximation even in the more general case (nonzero photon pumping and expansion) close to the line center and as most of the energy exchange occurs close to the line center, we use it for the discussion here. }.
\item[(c)]  For this setting, we  first assume the coupled \his-Lyman-$\alpha$ system to be in thermal equilibrium at $T= T_1$.  Then an energy 
$\Delta E$ (per neutral atom) is injected into the gas. If the gas was not coupled to Lyman-$\alpha$ photons, this 
energy injection would raise the temperature of the gas by $\Delta T = \Delta E/k$. However, for the coupled system,
this energy would be shared between the  two components, and a  new equilibrium temperature of the coupled system  would be smaller than the one without Lyman-$\alpha$ photons. 
\item[(d)] To compute the new equilibrium temperature  $T_2$, we first note that the  equilibrium spectrum of radiation is flatter close to the line center, as $T_2 > T_1$ or there are more photons on the blueward side of the line center as compared to the equilibrium at a smaller temperature. These photons are transferred from the redward to the blueward side by  inverse Compton scattering as the gas is heated. 
\item[(e)]The thermal energy of photons (close to the line center) is $\int h(\nu-\nu_\alpha) 4\pi J(x) dx $. This allows 
us to  compute the amount of energy absorbed by the Lyman-$\alpha$ photons in a small volume $\Delta V$: $\Delta E_\alpha = q n_\alpha(0) k(T_2-T_1) \Delta V $. 
Here, $n_\alpha(0)$ is the number density of Lyman-$\alpha$ photons at the line center, and $q$ is a number whose exact 
value depends upon the limit of integration; it is on the order of unity.  The gas absorbs the energy $\Delta E_g = n_b k(T_2-T_1) \Delta V$.  The new equilibrium temperature is computed by equating the injected energy with the energy absorbed by the two components:  $n_b   \Delta E  = (\Delta E_g + \Delta E_\alpha)/\Delta V $. This gives: $T_2 = T_1 +  n_b\Delta E/k(n_b + qn_\alpha(0))$.
\item[(f)] The heat capacities of the two components are proportional to their respective number densities. For $n_\alpha(0) \gg n_b$,  most of the thermal  energy is absorbed by Lyman-$\alpha$ photons, and the resultant temperature change is small. If $n_\alpha(0) \ll n_b$, most of the heat is absorbed by neutral atoms. 
\item[(g)] Figure~\ref{fig:heat} shows   solutions of  the general initial value problem: the Lyman-$\alpha$ photon injection and X-ray heating start at the same time and the expansion of the Universe is taken into account. However, the foregoing discussion also allows us to discern the main features of the dynamics of the system in this case.  For small $C$, the temperature of the gas keeps rising owing to 
continuous injection of energy from X-ray photons\footnote{It is instructive to compare  different time scales. The energy redistribution owing to inverse Compton scattering  is the  shortest. The X-ray heating time scale is shorter than the expansion time scale, which means the adiabatic cooling does not play an important role in Figure~\ref{fig:heat}.}.   In this limit, $n_\alpha(0) \lesssim n_b$ and the thermal evolution of the gas is nearly independent of the the presence of Lyman-$\alpha$ photons. However, owing to strong thermal coupling between the two components, the Lyman-$\alpha$ spectrum close to $x\simeq 0$ evolves rapidly 
to establish equilibria for different temperatures. As $C$ is increased, the temperature evolution flattens, reaches a peak, and then driven to a fixed temperature. This is owing to several factors, which are linked to the thermal evolution without any additional heating (Figure~\ref{fig:long_ns}) and,  for this range of $C$,  an  increasingly larger fraction of 
the heat is absorbed by Lyman-$\alpha$ photons.  For  even larger $C$, after initial heating, the system relaxes to a quasi-equilibrium temperature and remains close to
it during the later evolution of the system. This behaviour is nearly independent of the heating rate because these cases correspond
to $n_\alpha(0)/n_b \gg 1$ \footnote{In Figure~\ref{fig:long_noz}, $n_\alpha(0) \simeq 10 n_b$ for $C = 10^{-16}$ and $t \gtrsim 50 \rm Myr$. The figure also shows $n_\alpha(0)$ scales nearly linearly with $C$ and therefore $n_\alpha(0)/n_b \gg 1$ is achieved for most models of Lyman-$\alpha$ injection rates we consider in Figure~\ref{fig:heat}.}. 
\item[(h)] Finally, a comparison of  Figures~\ref{fig:long_ns} and~\ref{fig:heat} allows us to gauge the net impact of additional heating. For small $C$, the  thermal evolution for the gas with and without additional heating are quite different. This is expected based on our discussion.  As $C$ is increased, the thermal evolution of the gas begins to resemble the trajectory  without X-ray heating, 
and for large $C$, the thermal evolution is nearly independent of the  additional heating,  which is expected as $n_\alpha(0)/n_b \gg 1$ in these cases. 
\end{itemize}
In the next section, we explore this issue further for more realistic models of Lyman-$\alpha$ photon injection and X-ray heating.

\subsection{\texorpdfstring{Lyman-$\alpha$}{Lyman-alpha} and X-ray from Ionizing Sources}
In the foregoing, we considered Lyman-$\alpha$ injection and X-ray heating rates over a large range of values, but we did not consider cases where they arise from the same set of sources. This is our expectation in a more realistic setting. In the literature, multiple
models have been considered that simultaneously treat ionization, Lyman-$\alpha$ injection, and X-rays. 
In this paper, we explore the analytical models considered in \cite{RS19}. In these models, the excursion set formalism is used to calculate the collapse fraction ($f_{\rm coll}$) and the volume-averaged ionization fraction ($x_{\rm HII}$) as a function of the redshift \citep{FZH04a}. The same sources cause injection of X-ray and \lyal photons into the medium (e.g.,  \cite{RS18} and \cite{RS19} for details; while these papers considered the inhomogeneities of these radiation fields to compute the fluctuating component of \hi signal, we are only using volume-averaged quantities in this paper as our main interest is the global \hi signal). In this paper, our main focus is on the era of CD.
In this era, the ionization fraction is small, and the only relevant effects arise from \lyal and X-ray photons. 

The dynamics of the system is modelled using four parameters: the ionization efficiency factor $\zeta = 7.5$, the X-ray spectral index $\alpha = 1.5$, the total number of X-ray photons emitted per stellar baryon $N_{\rm heat}$, 
\footnote{Comparing the definition of $N_{\rm heat}$ with another parametrization of X-ray heating,  $\zeta_X$, the number of photons per solar mass in stars (used in the  publicly available code 21cmFAST \citep{21CMFAST}), we get $N_{\rm heat} \simeq \zeta_X m_p$. Hence, $N_{\rm heat} = 1$ would approximately correspond to $\zeta_X = 1.2 \pow{57} M_\odot^{-1}$.}
and the ratio of Lyman-$\alpha$ to hydrogen-ionizing photons emitted from the sources, $f_L$ \footnote{In the first part of the paper, we used a constant rate of Lyman-$\alpha$ photon injection, parametrized by $C$. It is not straightforward to relate $f_L$ to $C$ as the Lyman-$\alpha$ injection rate is time-dependent for a given $f_L$, which translates to a changing $C$. 
For $f_L = 10^4$,  $C = \{6\times 10^{-18}, 4 \times 10^{-17}, 6 \times 10^{-17}\}$  at redshifts $z = \{20, 15, 12\}$, respectively. }.

We initialize Eqs.~(\ref{eq:ts_def}), (\ref{eq:fineq2}), (\ref{eq:thermal}), and~(\ref{eq:ene_exchange1}) at $z=25$ and evolve Lyman-$\alpha$ profiles and the IGM temperature for 500~Myr, which means the simulations end at $z\simeq 8$.   We show these results in Figure~\ref{fig:g21}. We only vary $N_{\rm heat}$ and $f_L$ in displaying our results, as the  results 
are robust to changes in the other two parameters\footnote{While modelling the impact of Lyman-$\alpha$ and X-ray, it should be noted
that Lyman-$\alpha$ photons travel much further into the medium (a few hundred Mpcs, e.g., \cite{RS19} for details and further references).
For the purposes of this paper, we assume that all the Lyman-$\alpha$ photons emitted from the source are absorbed instantaneously; this gives us consistency with our notation in the earlier section. While
such an assumption could have a major impact on the fluctuating component of the \hi signal (e.g., \cite{RS19}), its main impact for our purposes
is to cause a sharper rise of \lyal  intensity initially, close to $z\simeq 25$, but the temperature evolution (and the global \hi signal) at lower redshifts is not affected.}.

We first analyze the case where the X-ray heating rate is low (lower left panel of Figure~\ref{fig:g21}). In this case, the net thermal impact of Lyman-$\alpha$ photons is to heat the system. This, as noted above, is expected and consistent with the results shown in Figure~\ref{fig:long_ns}. However,
in the middle and top left panels in which the X-ray heating rates are larger, the Lyman-$\alpha$ photons prevent heating. This is in line
with the discussion on Figure~\ref{fig:heat}. The most important result of the analysis, already anticipated in Figure~\ref{fig:heat}, is
that in all cases, for a large enough $f_L$, the temperature reaches an equilibrium value, which is close to the equilibrium temperature
reached without X-ray heating (Figure~\ref{fig:long_ns}). This equilibrium temperature is entirely determined by the ratio of injected to continuum photons (for details, see Paper~I).  In Figure~\ref{fig:g21}, we assume the fraction of injected photons to be 20\%, which is close to
the value expected from the quantum mechanics of the photon cascade and the spectrum of ionizing sources (e.g., \cite{chuzhoy2006ultraviolet,chuzhoy2007heating,furlanetto2006scattering,hirata2006wouthuysen}).  \citet{chuzhoy2007heating} argued that the ratio of injected to continuum photons depends on the temperature of ionizing sources. For hot sources, the ratio reaches 0.17, and it is expected to be greater than 0.1 for cooler sources. This motivates the range we consider in the next section (Figures~\ref{fig:g21} and~\ref{fig:g21_10}). 

There are two more contributions to injected photons that we have neglected in the paper. (a) Recombination photons from ionizing regions: In ionization equilibrium,  each ionizing photon causes a recombination that finally cascades to a 
Lyman-$\alpha$ photon  \footnote{The recombination time scale for redshifts of interest 
could be around  one-tenth of the  expansion time scale, so recombination is not that efficient. However, we will assume it here for our estimates}. This contribution is expected to be negligible for $f_L \gg 1$, (b) \hi excitations from X-ray photons: Through secondary excitations, a single X-ray photon can excite multiple hydrogen atoms. Up to 30\% of the energy of X-ray photons could be spent in exciting hydrogen atoms (e.g. \cite{1985ApJ...298..268S,Heating2001}). We have checked that this
contribution to the injected photon budget is also negligible for our models and it is generally smaller than the contribution from recombination photons. 

We can approximate  these contributions by neglecting expansion of the Universe, varying rates of different physical processes, etc.  If the ionized fraction of the IGM is  $f_{\rm ion}$, we expect the number density of hydrogen-ionizng photons, $n_{\rm ion} \simeq q n_b f_{\rm ion}$; here $q \simeq 10$ takes the impact of recombination into account. We know that nearly 20\% of X-ray photon energy goes to heat the mostly neutral medium. For simplicity, let us assume all X-ray photons have the same energy $E$. Nearly 30\% of the energy of this photon goes into 
the excitation of the hydrogen atom to the first excited state, or each X-ray photon produces $m = p E/E_\alpha$ injected photons; here $p =0.3$ and $E_\alpha = 10.2 \, \rm eV$ is the energy of the Lyman-$\alpha$ transition. If the medium is heated to a temperature $T$, we expect: $f n_x E \simeq (1-f_{\rm ion}) n_b kT_{\rm K} $; here $f = 0.2$ and $n_x$ is the number density of X-ray photons. Hydrogen-ionizing photons ionize regions close to ionizing sources.   The number of injected photons created by recombination in ionized regions ($N_{\rm ion}$) and by X-ray excitations in neutral regions ($N_{\rm neu}$) can now be computed. This gives us: $N_{\rm neu}/N_{\rm ion} \simeq (n_x m)/n_{\rm ion} \simeq (1-f_{\rm ion})p kT_{\rm K}/(f E_\alpha q f_{\rm ion})$.  Notice that this ratio is independent of the energy of X-ray photons.  Using numbers suitable for our study, $T_{\rm K} \lesssim 1000 \, \rm K$ and $f_{\rm ion} \simeq 0.1$, we can show that  $N_{\rm ion}/N_{\rm neu} \ll 1$.

As shown in Paper~I, the equilibrium temperature is higher/lower for a lower/higher fraction of injected photons. The constancy of the final
temperature irrespective of X-ray heating has important implications for the \hi signal from CD/EoR eras, which we discuss next. 

\subsection{Implications for the 21~cm Signal}

\begin{figure*}
  \includegraphics[width=1\linewidth]{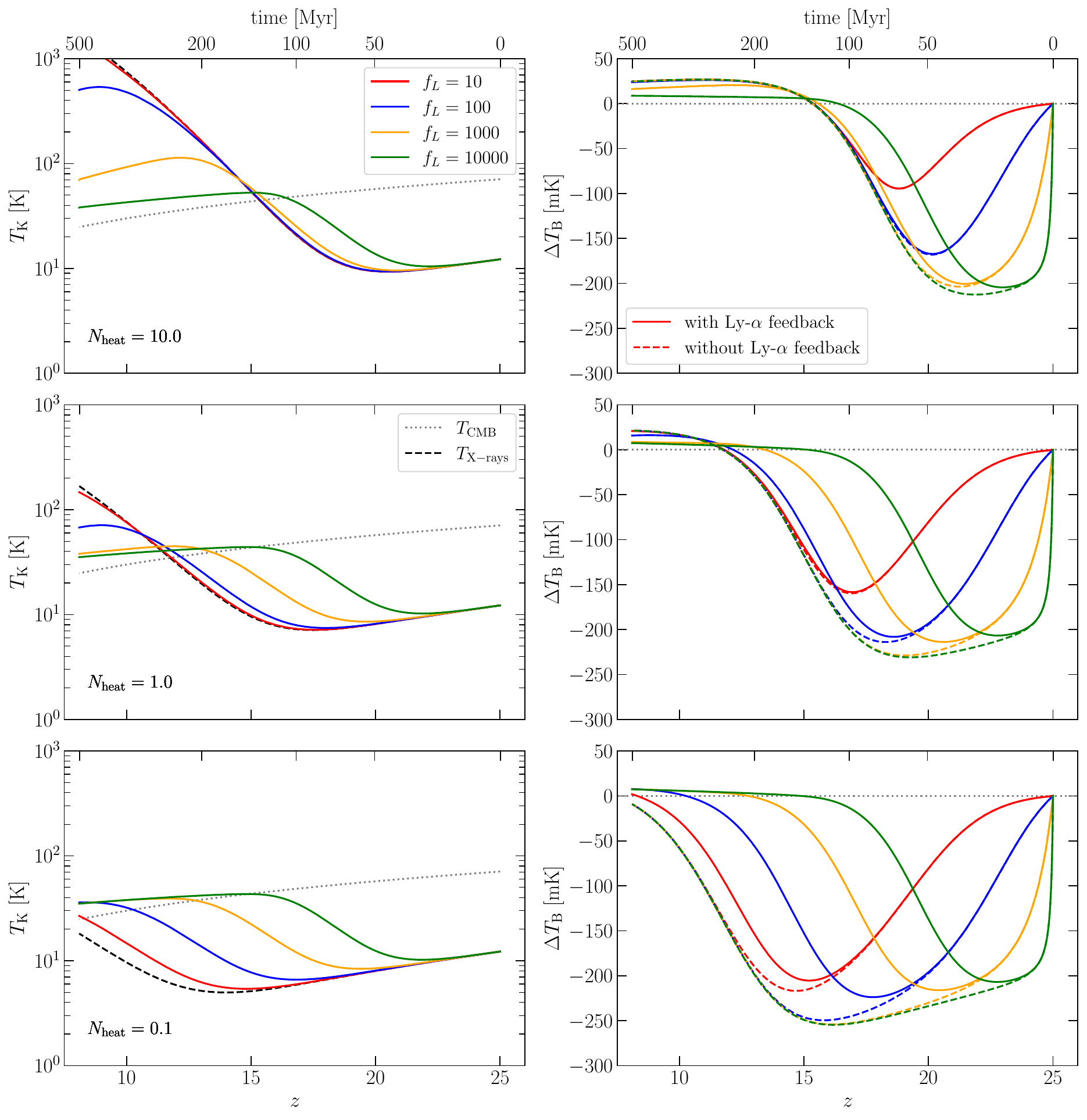} 
   \caption{We show the temperature evolution of the  gas  (left panels) and the global 21~cm signal (right panels) for the models parametrized by $N_{\rm heat}$ and $f_L$ (see the text for details). The Lyman-$\alpha$ model corresponds  to continuum + 17\% injected photons.  In each panel, we fix $N_{\rm heat}$ and vary $f_L$ by several orders of magnitude: $N_{\rm heat} = 10$ (top panel), 1 (middle panel), and 0.1 (bottom panel). {\it Left Panels}: Thermal history is displayed with (solid curves) and without (dashed curve) the thermal effects of  Lyman-$\alpha$ photons.  These results are in agreement with thermal histories shown in Figure~\ref{fig:heat}, even though the system  is evolved for 500~Myr here. {\it Right Panels}: The global \hi signal is displayed (Eq.~(\ref{overallnorm})). There are two curves (solid and dashed) for each model (fixed $f_L$ and $N_{\rm heat}$). For dashed curves, Wouthuysen-Field coupling  is included (Eq.~(\ref{eq:ts_def}) and the right panel of Figure~\ref{fig:long_ns})  but not the thermal impact of Lyman-$\alpha$ photons. Solid curves include all the relevant physics of Lyman-$\alpha$ photons. 
    For small values of $f_L$ (small number density of \lyal photons), there is negligible difference between solid and dashed curves. For higher values of $f_L$, the  inclusion  of \lyal thermal effect causes  the gas to heat faster initially. However,  \lyal photons prevent the gas temperature from  rising above the equilibrium temperature for the corresponding  Lyman-$\alpha$ model (continuum + 17\% injected photons), which is close to 40~K.}
  \label{fig:g21}
\end{figure*}

\begin{figure*}
  \includegraphics[width=1\linewidth]{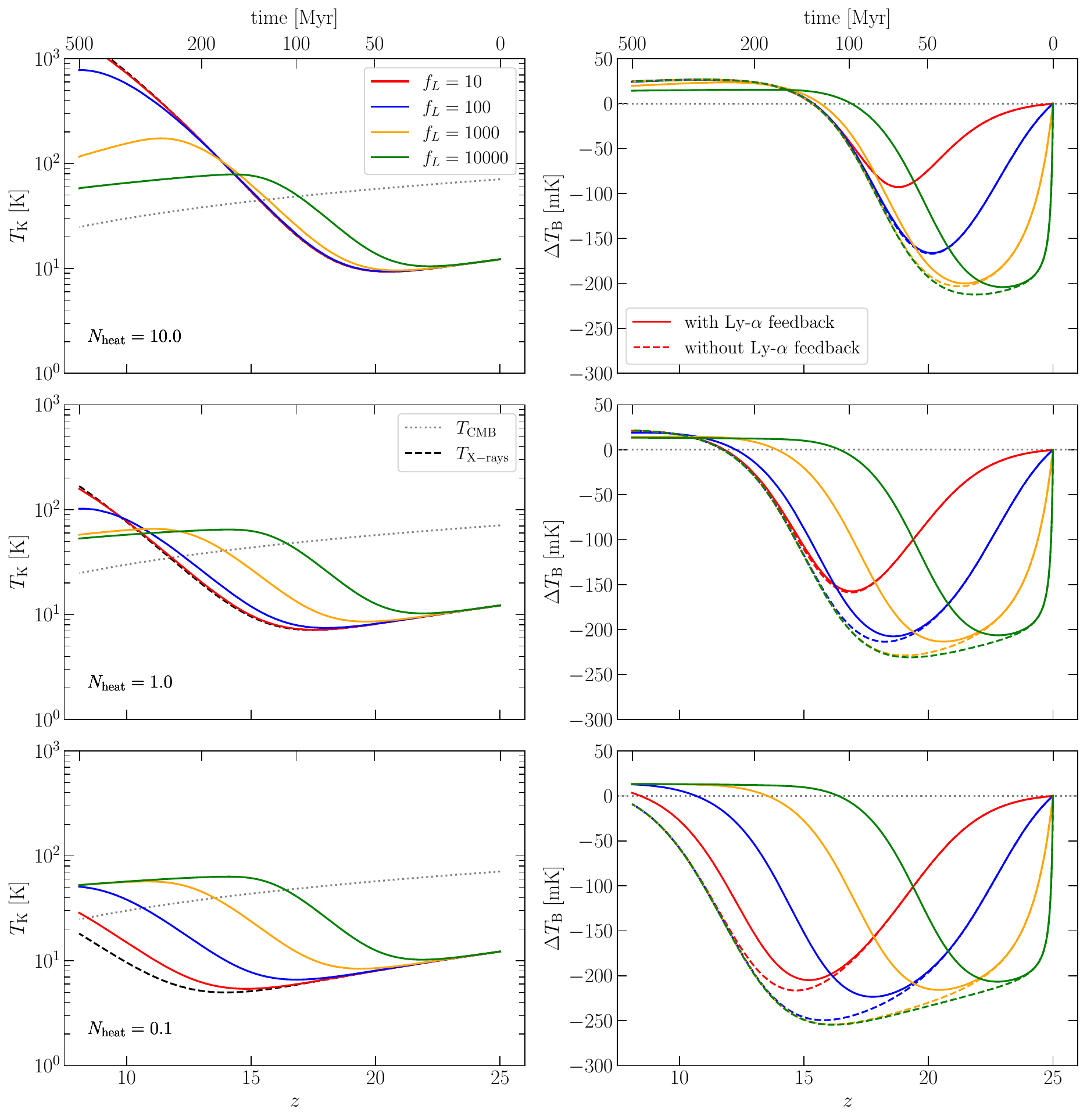} 
   \caption{The same as Figure~\ref{fig:g21} but for Lyman-$\alpha$ model, continuum plus 10\% injected photons. The equilibrium temperature in this case is $T \simeq 70 \, \rm K$, which gives the redshift of crossover of the global  \hi signal from absorption to emission, $z_c \simeq 17.5$.} 
  \label{fig:g21_10}
\end{figure*}

The observable global  \hi signal, the difference of \hi brightness  temperature and CMB temperature, can be expressed as  (e.g., \cite{21cm_21cen,1997ApJ...475..429M,1999A&A...345..380S,2004ApJ...608..611G,Sethi05, RS19}): 
		\begin{equation}
				\Delta T_b	 \simeq 26.25 x_{\rm \scriptscriptstyle HI} \left(1-\frac{T_{\text{CMB}}}{T_{\rm S}}\right) \left(\frac{1+z}{10} \right )^{1/2} \left(\frac{0.14}{\Omega_m h^2}\right)^{1/2} \left(\frac{\Omega_b h^2}{0.022}\right) \text{mK} \label{overallnorm}
	\end{equation}
This signal is determined by the evolution of spin temperature, $T_{\rm S}$, which is given by  Eq.~(\ref{eq:ts_def}); $T_{\rm S}$ is
determined simultaneously with Eqs.~(\ref{eq:fineq2}), (\ref{eq:thermal}), and~(\ref{eq:ene_exchange1}). The system of equations
is initialized at $z=25$ with the matter temperature $T_{\rm K} \simeq 10 \, \rm K$ (left panels of Figure~\ref{fig:g21}), which corresponds to  the  post-recombination,  unheated gas (for details, see, e.g., \cite{RS18} and references therein). The global
\hi signal, corresponding to the thermal history given in the left panels, is displayed in the right panels of Figure~\ref{fig:g21}. The lower
right panel shows the case when X-ray heating is small. In this case, as already noted, the net impact of Lyman-$\alpha$ photons is to provide heating in addition to X-ray photons, which,  it follows from Eq.~(\ref{overallnorm}), means the \hi signal is shallower in absorption. In the other two cases, the net impact of Lyman-$\alpha$ photons is to provide initial heating followed by cooling. 

The most striking feature of the \hi
signal in all three panels occurs in the limit when Lyman-$\alpha$ photon density becomes large. In this case, as discussed above, the system relaxes to
an equilibrium temperature, which is independent of both the Lyman-$\alpha$ photon density and the heating rate. As $T_\alpha \simeq T_{\rm K}$ and $y_\alpha \gg 1$ (right panel of Figure~\ref{fig:long_ns}), from Eq.~(\ref{eq:ts_def}) it follows that $T_{\rm S} \simeq T_{\rm K}$. This means that for a given neutral fraction, $x_{\rm \scriptscriptstyle HI}$, the global \hi signal is frozen to a pattern that is independent of both Lyman-$\alpha$ photon intensity and the X-ray heating rate. During the cosmic dawn, $x_{\rm \scriptscriptstyle HI} \simeq 1 $, and Eq.~(\ref{overallnorm}) shows that the  global \hi signal is entirely determined by $T_{\rm S}$ during this era. As $T_{\rm S}$ is known in the limit of large Lyman-$\alpha$ photon density, the redshift, $z_c$, at which the global \hi signal makes a transition from absorption to emission is also fixed. For the injected to the continuum ratio we consider in Figure~\ref{fig:g21},  $T_{\rm S} \simeq 40 \, \rm K$, which, from Eq.~\ref{overallnorm}, gives $z_c \simeq 13$. During the EoR, $x_{\rm \scriptscriptstyle HI}$ 
varies substantially and finally vanishes.  Therefore, by fixing $T_{\rm S}$, we cannot determine $\Delta T_b$ during this era. However, as $x_{\rm \scriptscriptstyle HI} < 1$, we obtain an upper limit on the \hi signal. In Figure~\ref{fig:g21_10} we consider
a different model of Lyman-$\alpha$ injection (continuum plus 10\% injected photons). The most notable difference between
Figures~\ref{fig:g21} and~\ref{fig:g21_10} is the equilibrium temperature for a large Lyman-$\alpha$ photon density. As expected and shown
in Paper~I, the equilibrium temperature is higher for a smaller fraction of injected photons. In the case shown in Figure~\ref{fig:g21_10},
the equilibrium temperature is close to 70~K, which gives $z_c \simeq 17.5$. All features of the global \hi signal shift to higher redshifts in this case. 

The previous paragraph summarizes the most important outcome of our analysis. Our study shows that, for high Lyman-$\alpha$ injection rates, the global \hi signal is greatly simplified. Figure~\ref{fig:g21} further demonstrates that the thermal impacts of Lyman-$\alpha$ and X-ray photons are coupled and treating them
as uncoupled photon fields, as is usually the case, does not treat the effect of either adequately. 

Generally,  the \hi global signal is more complicated. Figure~\ref{overallnorm} shows that for intermediate values of
photon injection rates, the \hi signal is not fully fixed. Here, the main impact of Lyman-$\alpha$ photons is still to
drive the gas temperature to the equilibrium temperature, which causes a decrease in the \hi signal in both absorption and emission. In such cases, it is not possible to fix the redshift of crossover from absorption to emission. 
 For lower Lyman-$\alpha$ injection
rates  ($C \lesssim 10^{-18}$),  the thermal impact of Lyman-$\alpha$ photons is negligible, and the Lyman-$\alpha$--X-ray system can be treated as uncoupled (Figures~\ref{fig:heat} and~\ref{fig:g21}). However, even in this case, it is possible to couple the \hi spin temperature to the kinetic temperature (this requires $y_\alpha \gg 1$, Eq.~(\ref{eq:ts_def})) on time scales shorter than the expansion time scale (right panel of Figure~\ref{fig:long_noz}). This shows that a parameter space exists in  which Lyman-$\alpha$ photons can play their
traditional role, $T_{\rm S}\hbox{--}T_{\rm K}$ coupling, without any significant thermal impact. 

\section{Summary and Conclusions} \label{sec:conc}
In this paper, we extend our work from Paper~I, which was mainly a theoretical study.  In Paper~I, we solved  the thermal evolution of the coupled Lyman-$\alpha$--\hi gas system as an initial value problem.  In this paper, we carry out necessary improvements to our earlier analysis and demonstrate the relevance of our work for the \hi signal during CD.

Our main results can be summarized as follows: 
\begin{itemize}
\item[(1)] {\it Long-term evolution of photon profile and gas temperature}: The photon profile stops evolving on long enough time scales  if  the expansion of the Universe is  neglected. The time-scale over which this equilibrium is reached depends on $x_{\rm max}$ and is around 50~million years for the realistic case considered in this paper. With the inclusion of expansion, the photon profile evolves more rapidly during the flat profile phase   but slowly after the tilted profile phase is reached (Figure~\ref{fig:long_noz}). For more details on the transition from flat to tilted profile and approach to equilibrium, see Paper~I. 
The long-term evolution of gas temperature shows that, when the expansion is neglected,   an equilibrium is reached on  a  time scale that varies inversely with photon injection rate $C$ (left panel of Figure~\ref{fig:long_ns}). As shown in Paper~I, the equilibrium temperature
is independent of the photon injection rate and depends only on the ratio of injected to continuum photons. The long-term thermal evolution shows a slow decrease at $t \gtrsim 10 \, \rm Myr$ owing to the expansion of the Universe. 
\item[(2)] {\it X-ray as additional heating source}: We expect X-ray from accretion around the first sources to heat the neutral IGM around ionized regions 
during CD/EoR eras. We include this effect in addition to the thermal impact of Lyman-$\alpha$. In the literature, these two are treated as
uncoupled. However, our analysis shows that Lyman-$\alpha$ photons--\hi gas--X-ray photons constitute a strongly coupled system in the limit 
when the Lyman-$\alpha$ injection rates are large. As shown in Figure~\ref{fig:heat}, both Lyman-$\alpha$ and X-ray photons heat at early times, so they act as uncoupled sources of heat. This behaviour is also seen when  the injection rate is low ($C =  10^{-18}$). At longer times and large injection rates, the next impact of Lyman-$\alpha$ photons is to prevent X-ray heating. For large $C$, the system is 
driven toward an equilibrium temperature that is solely determined by Lyman-$\alpha$ photons (Figure~\ref{fig:long_ns}). Our result
underlines the importance of  treating the issue as an initial value problem and as a coupled system. While our analysis suggests a major
rethink on how this coupled system is treated, it is not needed for small injection rates,  $C \lesssim 10^{-18}$. In this case, Lyman-$\alpha$ photons make a negligible thermal impact (Figure~\ref{fig:heat}) and are not coupled to X-ray heating (Figure~\ref{fig:g21}).
However, even in this case, the Wouthuysen-Field effect efficiently couples the hyperfine \hi level to the gas temperature as one reaches $y_\alpha \gg 1$ over  time scales shorter than the expansion time scales (right panel of Figure~\ref{fig:long_ns}). 
\item[(3)] {\it \hi signal}:  For computing the \hi signal, we consider specific reionization models in which ionizing, Lyman-$\alpha$, and 
X-ray photons are emitted by the same sources. The thermal histories and the global \hi signal for these models are shown
in Figure~\ref{fig:g21}. For a small Lyman-$\alpha$ injection rate, Lyman-$\alpha$ acts as an independent heating source, which makes 
the  \hi signal during the CD era shallower. The distinguishing feature of our analysis  arises again when the Lyman-$\alpha$ injection is large. As the gas temperature gets frozen in this case, this 
completely determines: (a) the \hi signal   during the CD era when $x_{\rm HI} \simeq 1$ and (b) the crossover redshift  from absorption to emission. These features depend  only on   the  ratio of injected to continuum photons.  This is the most important result of our analysis. 
\end{itemize}

In our analysis, we assume two models of Lyman-$\alpha$ injection: (a) a source with a constant injection rate $C$  (Figure~\ref{fig:heat}), and (b) a model based on sources of ionization that switch on
at $z = 25$ (Figures~\ref{fig:g21} and~\ref{fig:g21_10}). How robust are our results to more complicated evolution histories of Lyman-$\alpha$ injection? As discussed above and also in Paper~I, our results are mainly determined by the injection rate of Lyman-$\alpha$ photons and their number density and are therefore relatively insensitive to other complications. One case in which our results might radically alter is if the Lyman-$\alpha$ source is turned off. This causes
the medium to heat for the following reason: injected photons, which are responsible for cooling, redshift away from the resonance on short time scales ($t \lesssim 0.1 \, \rm Myr$, see Paper~I for details) while continuum photons continue redshifting into
the resonance from the blueward side for $t \simeq 100 \, \rm Myr$. The continuum photons, even as their number gets diluted owing to the change in redshift, continue heating the medium over this
time scale before the tilted profile is reached (Figure~\ref{fig:long_noz} and Paper~I).  While it is possible to build
models in which ionizing sources are switched off, e.g., if the haloes responsible for ionization are molecular hydrogen-cooled
and they are destroyed by Werner band photons from these sources (e.g., \cite{Barkana:2000fd} and references therein), we expect the 
atomic-cooled  haloes to dominate the reionization process for $z \lesssim 15\hbox{--}20$ and their number is expected to increase. 
In summation,  our model captures the essence of relevant processes needed to understand the physical setting we consider. 

There are multiple ongoing experiments to detect the global \hi signal (EDGES, \cite{2018Natur.555...67B};  SARAS, \cite{2022NatAs...6..607S}; REACH, \cite{2022NatAs...6..984D};  LEDA, \cite{2018MNRAS.478.4193P}; PRIzM, \cite{2019JAI.....850004P}; MIST, \cite{2024MNRAS.530.4125M}). 
Most of these experiments target a
range of frequencies that cover the redshift range $7 <z < 20$. Generally, these experiments search through thousands of global \hi signal templates to determine
the parametrized theoretical model.  Figures~\ref{fig:g21} and~\ref{fig:g21_10} capture the diversity of the global \hi signal.  However, as already noted above, our work makes a definitive prediction if  the Lyman-$\alpha$ photon density reaches large values  at early times. In this case,   the temperature of the medium becomes nearly independent of the heating rate. Figures~\ref{fig:g21}  and~\ref{fig:g21_10}  partially capture  this limit but it is seen more clearly in Figure~\ref{fig:heat} for larger values of $C$.
In this limit, the global \hi signal can be determined during the era of CD. Such a signal  would be easier to search for, e.g., in terms of three
parameters: the crossover redshift $z_c$, the depth of the signal for a fixed redshift $z > z_c$, and the height of the signal for a fixed redshift $z < z_c$.
For injected-to-continuum photon ratio in the range between 0.10 and 0.17, as expected from theory (e.g., \cite{chuzhoy2007heating}), $17.5 < z_c < 13$ (Figures~\ref{fig:g21} and~\ref{fig:g21_10}). This means that the global \hi signal might be able to  determine this ratio, which  depends on
the quantum mechanics of the hydrogen atom and some properties of the ionizing sources.


\section*{Acknowledgments}
We would like to thank the anonymous referee for useful comments and suggestions, which helped us improve this paper.

\bibliographystyle{aasjournal}
\end{document}